\documentclass[11pt,a4paper]{llncs}
\usepackage{amsmath}
\usepackage{amssymb}
\usepackage{graphicx}
\usepackage{marvosym}
\usepackage{url}
\usepackage{fancyhdr}
\usepackage[utf8]{inputenc}
\usepackage{listings}
\usepackage{xcolor}

\usepackage{geometry}
\newcommand{\keywords}[1]{\par\addvspace\baselineskip
\noindent\keywordname\enspace\ignorespaces#1}

\lstdefinelanguage{alloy}{
	morekeywords={
		module, open, as,
		private, abstract, sig, extends, in,
		lone, some, one, disj,
		fact, pred, fun, assert,
		run, check,
		for, but, exactly,
		this, not, implies, else, let,
		not, no, set, all, sum,
		iff, or, Int, and,
		none, univ, iden
	},
	breaklines=true,
    framesep=12pt,
    showstringspaces=false,
    tabsize=4,
    captionpos=b,
    showstringspaces=false,
	basicstyle=\ttfamily\small,
	morecomment=[l]{//},
	morecomment=[l]{--},
	morecomment=[s]{/*}{*/},
	morestring=[b]{"},
	commentstyle=\color[HTML]{117733}\itshape,
  	keywordstyle=\color[HTML]{332288}\bfseries,
  	ndkeywordstyle=\bfseries,
}[keywords,comments,strings]

\lstdefinestyle{alloy}{
	commentstyle=\itshape,
	keywordstyle=\bfseries,
	stringstyle=\itshape,
}

\usepackage[%
    colorlinks=false
]{hyperref}

\DeclareUnicodeCharacter{FB01}{fi}

\fancypagestyle{firstpage}{\fancyhf{}}

\begin{document}
\title{\LARGE{Formal Verification of Ecosystem Restoration Requirements using UML and Alloy}}

\author{\large{Tiago Sousa  \and Benoît Ries \and Nicolas Guelfi}}
\institute{\large{Department of Computer Science, University of Luxembourg,\\ Esch-sur-Alzette, Luxembourg}}
\maketitle

\thispagestyle{firstpage}

\begin{abstract}
United Nations have declared the current decade (2021-2030) as the "UN Decade on Ecosystem Restoration" to join R\&D forces to fight against the ongoing environmental crisis. 
Given the ongoing degradation of earth ecosystems and the related crucial services that they offer to the human society, ecosystem restoration has become a major society-critical issue. It is required to develop rigorously software applications managing ecosystem restoration. Reliable models of ecosystems and restoration goals are necessary.
This paper proposes a rigorous approach for ecosystem requirements modeling using formal methods from a model-driven software engineering point of view. The authors describe the main concepts at stake with a metamodel in UML and introduce a formalization of this metamodel in Alloy. The formal model is executed with Alloy Analyzer, and safety and liveness properties are checked against it. This approach helps ensuring that ecosystem specifications are reliable and that the specified ecosystem meets the desired restoration goals, seen in our approach as liveness and safety properties.
The concepts and activities of the approach are illustrated with CRESTO, a real-world running example of a restored Costa Rican ecosystem.
\end{abstract}

\section{Introduction}

Ecosystems are complex systems that are crucial to the sustainability of human life on Earth. With the growing threat of climate change and environmental degradation, there is an increasing need for software applications that can help us understand and manage the restoration of ecosystems.  Initially the concept of "Ecosystem restoration" was restricted to bringing back a degraded ecosystem to its previous state~\cite{gannInternationalPrinciplesStandards2019}. Nowadays, this concept has been loosened to integrate climate change and the possibility that strictly reversing to previous state might not be the optimal solution, UN now defines Ecosystem restoration as “the process of halting and reversing degradation, resulting in improved ecosystem services and recovered biodiversity.[...]"~\cite{unitednationsenvironmentprogrammeunepBecomingGenerationRestorationEcosystem2021}.

However, developing software applications dealing with ecosystems data is a challenging task, as it requires a deep understanding of the ecosystem's structure, behavior, and its interactions with its environment. During the last decades, the software engineering community has tackled the problem of modeling complex systems and the verification of properties on these models. In this paper, we focus our contribution on two early activities of the software development lifecycle, namely requirements modeling and formal verification of requirements models. On the one hand, requirements modeling is typically used in Software Engineering \cite{sommervilleSoftwareEngineering2016} to improve the communication between stakeholders before designing and actually producing the system. On the other hand, formal verification techniques are used in different phases of the software development lifecycle. Our approach focuses on using formal verification to ensure that the ecosystem requirements model meet some specified restoration-related properties.

In the context of natural ecosystem restoration, the duration for these restoration processes may take from years to decades before seeing the actual results. As raised by the UN, there is an urgency to deal correctly and rapidly with the issues of the current situation both in terms of financial impact of degraded ecosystems to our society and also in terms of human impact of the sustainability of our society. That is why we consider software applications dealing with ecosystem restoration as critical-software applications. As such, formal methods are necessary to guarantee a rigorous requirements modeling.

The aim of this paper is to show the usefulness in a short-term time-frame of formal methods for ecosystem restoration requirements. Due to the aforementioned pressuring financial and human issues, our approach aims at increasing confidence in requirements model and provide guarantee on the properties of these models.

Our approach is designed in the context of Model-Driven Engineering~\cite{kentModelDrivenEngineering2002,bezivin04}. MDE places models as first-class citizens within software development lifecycles. Meta-modeling is a key activity in MDE approaches. Meta-models define a modeling language tailored to specific concepts of an application domain. Moreover, the definition and usage of a metamodel in an MDE approach allows for \emph{model transformations}, from one modeling language to another modeling language, e.g., from a requirements model to a design model.

Following a typical software engineering approach, requirements artifacts are used as input for the further development activities, i.e. design and production of the software applications~\cite{sommervilleSoftwareEngineering2016}. In our context, the intention of producing these formally verified requirements models of ecosystems is to use them as input for the development of software applications helping to manage and understand restoration challenges of ecosystems~\cite{sousaModelingPredictingResilience2022,sousaEcosystemResilienceAnalysis2023}.

Future work includes the development of a simulation software application that takes this verified ecosystem restoration requirements model as input for its simulation configuration. Another future work thereafter is the development of an AI-based software application that generates synthetic data based on the verified ecosystem restoration requirements model provided by this approach, in order to fill the potential gaps of missing real-world data.

In this paper, our main contribution is the proposal of a formal approach for modeling and verifying the requirements of ecosystems restoration goals using the standardized Unified Modelling Language (UML) and the Alloy formal language. We illustrate our approach with a case study by Treuer et al.\cite{treuerLowcostAgriculturalWaste2018}  on the regeneration of a Costa Rican ecosystem.

As a summary, our proposed approach provides tools for producing verified ecosystem requirements models including restoration goals. Our approach has the potential to enhance the rigor and reliability of requirements engineering and verification in the domain of natural ecosystems restoration. It can help ensure that the ecosystems specifications are reliable and meet some given desired restoration goals. We believe that our work makes a valuable contribution to the field of formal software engineering applied to ecosystems restoration and  have practical implications for researchers and practitioners, like political and environmental policymakers. As a further step, the verified ecosystem requirements model will be used in an MDE process serving as input for the development of AI-based and remote sensing-based applications for dealing with ecosystem restoration data.

The remainder of this paper is structured as follows. In Section 2, we present the running example of the Costa Rican ecosystem. We then provide, in Section 3, an overview of the requirements specification phase for ecosystem applications, combining it with a literature review. Section 3 also presents our UML metamodel for ecosystem requirements and describe the key concepts in natural language. We then introduce our formal verification approach, in Section 4, including a formal definition of Alloy concepts and a detailed description of the formal modeling process. Then, in Section 5, we review related work on the use of formal methods for requirements engineering and verification in the domain of natural ecosystems. We focus on recent studies that use UML and Alloy or similar formal methods, as well as studies that apply these methods to remote sensing applications. Finally, in Section 6, we discuss future directions for our research, how the contribution in this paper fits in a software development process.

\section{CRESTO Running Example}\label{section-runningexample}
In this section, we introduce the CRESTO\footnote{\textit{CRESTO} stands for "Costa Rica Restoration"} running example based on the real-life case study as presented by Treuer et al. \cite{treuerLowcostAgriculturalWaste2018}.

The study, conducted in the Guanacaste Conservation Area (GCA), a protected region in northwestern Costa Rica, addresses the pressing issue of tropical forest restoration. The GCA covers 1,000 square kilometers and encompasses a diverse range of habitats, including tropical dry forests affected by deforestation and degradation from human activities. The research in \cite{treuerLowcostAgriculturalWaste2018} focuses on exploring new methods for restoring these ecosystems, with particular attention given to the use of agricultural waste, to improve forest restoration. The study was carried out in two distinct zones within the GCA, designated as Modulo II experimental zone and Control Zone. Modulo II, a 3-hectare region, was treated with 12,000 metric tons of orange waste from a nearby juice factory, while the Control Zone, a neighbouring region of equal size, was left untreated. Both zones had similar initial ecological characteristics, referring to the variety of plant and animal species, vegetation, and soil composition, as well as initial degradation levels, which indicates the extent of damage to the natural ecosystem caused by factors such as pollution, deforestation or invasive species. Researchers monitored both zones for over a decade, collecting data on vegetation, soil, and overall ecosystem health. Their findings indicated significant improvements in Modulo II, such as plant diversity, soil fertility,  and more rapid forest restoration compared to the untreated Control Zone. This suggests that using agricultural waste can be an effective and inexpensive approach to restoring degraded tropical forests, which play a critical role in preserving global biodiversity and mitigating climate change.

We use CRESTO to illustrate the applicability of our approach to modeling ecosystem restoration requirements and checking safety and liveness properties by applying formal verification techniques.

\section{Requirements Specification of Ecosystems} \label{req-section}
In this section, we focus on the initial stage of a simple waterfall software engineering life cycle \cite{sommervilleSoftwareEngineering2016}: the requirements engineering phase. We delve into its application in the context of ecosystem restoration requirements. 

\begin{figure}[ht!]
	\centering
	\includegraphics[width=\textwidth]{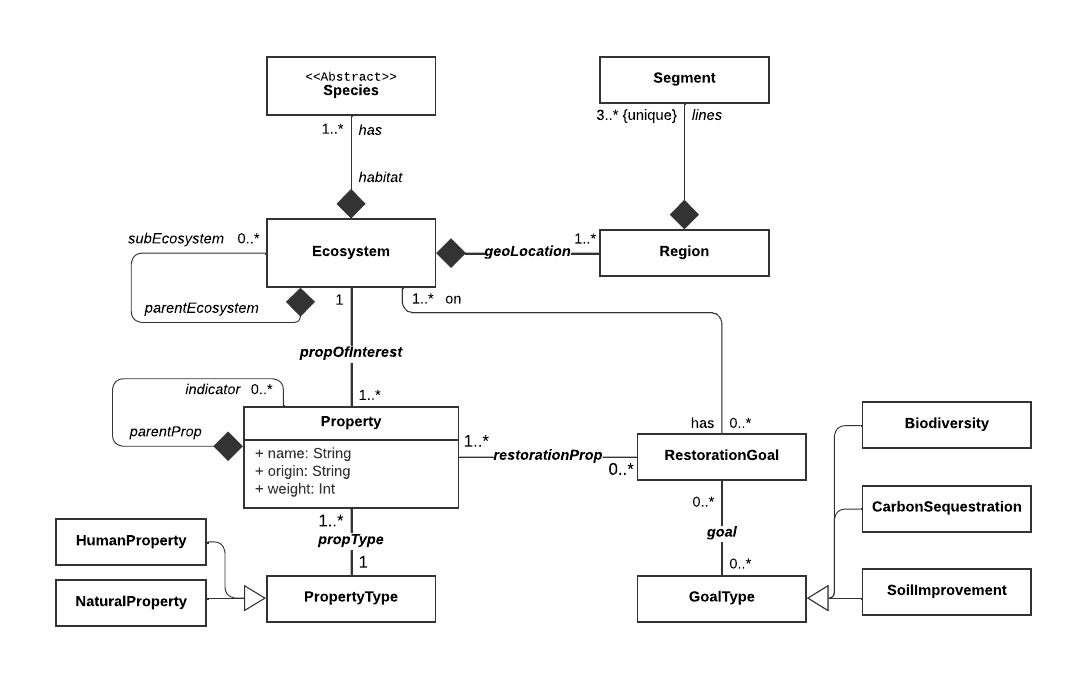}
	\caption{Metamodel for Ecosystem Requirements Specification}
	\label{fig:metamodel-uml}
\end{figure}

Understanding and managing intricate relationships between various entities and properties is important for the success of rehabilitation efforts \cite{gannInternationalPrinciplesStandards2019} such as restoration or monitoring of complex ecosystems. The specialized metamodel depicted in Figure \ref{fig:metamodel-uml} provides a semi-formal framework to capture and analyze these relationships, enabling stakeholders to make informed decisions related to ecosystem restoration requirements.

By presenting the different concepts within our specialized metamodel and emphasizing the motivation behind them, we aim to provide a comprehensive abstraction of the relationships and factors involved in ecosystem requirements. In the following subsections, we introduce and describe in a systematic manner each concept related to our metamodel, depicted in Figure \ref{fig:metamodel-uml}. First, each concept is presented informally, then we present how it has been integrated in our metamodel and finally, we illustrate its usage in CRESTO.

\subsection{Ecosystem}
In our approach, an ecosystem is a natural environment with its own unique characteristics, consisting of various species, regions, restoration goals, and potentially nested sub-ecosystems. It aims to capture the organization, complexity and interrelationships of the components within a natural environment and their sub-ecosystems.

In our metamodel (see Fig. \ref{fig:metamodel-uml}), the \texttt{Ecosystem} class has associations with \texttt{Species}, \texttt{Region}, \texttt{RestorationGoal}, and a recursive association with itself for \texttt{subEcosystems}. These relationships help model the different aspects of an ecosystem and its interactions with various components.

Finally, in CRESTO, the Guanacaste Conservation Area (GCA) is modeled as an ecosystem, as it is a protected region in northwestern Costa Rica where the study focuses on restoring tropical forests. 

\subsection{Region}
A region represents a geographical area of an ecosystem, defined by a set of line segments. It helps to model and define specific spatial aspects and characteristics of the environment.

In our metamodel (see Fig. \ref{fig:metamodel-uml}), the \texttt{Region} class has an association with the \texttt{Segment} class. This relationship helps define the boundaries of a region and allows for the modeling of complex shapes and areas within an ecosystem.

In CRESTO, both the Modulo II and the Control Zone are regions part of the GCA ecosystem. 

\subsection{Segment}
A segment is a line that helps define the boundaries of a region. It consists of start and end points with their respective X and Y coordinates, allowing for the accurate modeling of the shape and size of an ecosystem region.

The \texttt{Segment} class in our metamodel (see Fig. \ref{fig:metamodel-uml}) is used in a composition relationship with the \texttt{Region} class, enabling the construction of complex shapes and areas to accurately represent the different region zones of an ecosystem.

In CRESTO, the borders of the Modulo II and Control Zone regions are defined with a set of segments, helping to specify the area where the orange waste is applied and the area left untreated for comparison.

\subsection{Species}
A species is an abstract representation of living organisms that inhabit in one or more ecosystems. It models the diversity and interactions of various life forms within an ecosystem.

In our metamodel (see Fig. \ref{fig:metamodel-uml}), the \texttt{Species} class has an association with the \texttt{Ecosystem} class, indicating that a species inhabits an ecosystem. This relationship models the presence and distribution of different species as part of ecosystems.

In CRESTO, \texttt{Species} could represent various plants, animals, and microorganisms that are part of the tropical forest ecosystem in the GCA. For instance, the Hyparrehenia and Byrsonima plants are particular examples of a species in CRESTO. Hyparrehenia is a plant species found in the ModuloII region, whereas Byrsonima is a plant associated with the Control Zone. The diverse species found in these two regions exemplify restoration-related properties.

\subsection{RestorationGoal}
A restoration goal refers to one objective for restoring a specific ecosystem. Such goals can involve various objectives such as improving water quality or accelerate forest regeneration, to name a few.

In our metamodel (see Fig. \ref{fig:metamodel-uml}), the \texttt{RestorationGoal} class has associations with the \texttt{Ecosystem} and \texttt{Property} classes. These relationships models the context and conditions for achieving a restoration goal. Moreover, it has an association with the \texttt{GoalType} class, enabling the specification of different goal types for the restoration, such as biodiversity, carbon sequestration or soil improvement.

In CRESTO, the desired increase in biodiversity, soil fertility, and forest restoration with the application of the orange waste property in the Modulo II region is modeled as a \texttt{RestorationGoal}.

\subsection{Property}

The Property concept represents key factors, conditions, or characteristics associated with an ecosystem that play a crucial role in understanding its health, restoration, or degradation. By capturing these properties, stakeholders can gain valuable insights into the ecosystem's current state, assess the impact of specific interventions, and devise targeted strategies for restoration, conservation, or management, among others.

A property is a representation of various attributes and characteristics of an ecosystem that play a role in determining its health, restoration, or degradation. Properties are associated with a particular origin and a type, which helps classify them based on their nature and function in the ecosystem.

In our metamodel (see Fig. \ref{fig:metamodel-uml}), the \texttt{Property} class is associated with the \texttt{RestorationGoal} and \texttt{PropertyType} classes. These relationships represent the various types of properties and how they relate to the restoration goal of an ecosystem.

In CRESTO, the \texttt{Property} concept is used to represent the application of agricultural waste, specifically orange waste, in the restoration goal of the Modulo II zone. The waste's origin, from a juice factory in Costa Rica, and its classification as a human-originated factor, are important pieces of information to understand its effect on the ecosystem.

\subsection{PropertyType}
A property type serves as a means to classify various properties of an ecosystem. It differentiates between the distinct origins, nature, and functions of properties that influence the ecosystem's health, restoration, or degradation. Furthermore, it contributes to a comprehensive understanding of the ecosystem's characteristics and their impact on its restoration goals.

In our UML metamodel (see Fig. \ref{fig:metamodel-uml}), the \texttt{PropertyType} class has two different specializations, namely \texttt{HumanProperty} and \texttt{NaturalProperty}. Properties caused by human activities are specified with the \texttt{HumanProperty} specialization, while properties resulting from non-human factors are specified with \texttt{NaturalProperty}.

For instance, in CRESTO, the \texttt{PropertyType} is used to classify the type of orange waste as \texttt{HumanProperty} since it is sourced from a factory in Costa Rica.

\subsection{GoalType}
A goal type functions as a medium to classify restoration goals within an ecosystem. It categorizes restoration goals according to their specific objectives and targets, which are essential for understanding the desired outcomes of restoration efforts and measuring their progress.

In our UML metamodel (see Fig. \ref{fig:metamodel-uml}), the \texttt{GoalType} class has specializations named \texttt{Biodiversity}, \texttt{CarbonSequestration} and \texttt{SoilImprovement}. These specializations represent different aspects of ecosystem restoration that may be prioritized depending on the context and requirements of a particular ecosystem. By incorporating the \texttt{GoalType} concept and its specializations into the metamodel, stakeholders can better define, understand, and manage the diverse objectives of ecosystem restoration efforts.

In CRESTO, the desired increase in biodiversity, soil fertility, and rapid forest restoration with the application of orange waste in the Modulo II region is an example of a restoration goal that could be classified as \texttt{Biodiversity} and \texttt{SoilImprovement}.
 
\section{Formal Verification of Ecosystem Requirements} \label{alloy-section}
The previous section provides an overview of our specialized metamodel for ecosystem requirements specification, which captures and analyzes relationships between various entities and properties within ecosystems. The metamodel is designed to facilitate decision-making in ecosystem restoration requirements, while providing a semi-formal framework for modeling and understanding the concepts and relationships involved in ecosystem requirements. It presents the concepts in a systematic manner, and demonstrates how they have been integrated into the metamodel and modeled in CRESTO.

In this section, the focus shifts to the field of formal methods and their usage for the formal verification of ecosystem requirements within a software engineering context. As software systems become more complex, the need for robust verification and validation techniques grows increasingly important. Formal methods offer a rigorous, mathematically-based approach to specify and to verify software systems, thus increasing confidence in the validity of the requirements.

Hence, to create our formalization, we use Alloy \cite{jacksonAlloyLightweightObject2002}, a formal modeling and specification language based on first-order logic and inspired by the Z specification language \cite{spiveyNotationReferenceManual1992} and object-oriented modeling languages such as UML \cite{cookUnifiedModelingLanguage2017}. Alloy offers a powerful and expressive means to model and analyze complex requirements.

Our formalization is designed to allow stakeholders to accurately define the desired restoration requirements of an ecosystem, facilitating the detection, the elimination of inconsistencies and errors in the early stages of the software development lifecycle. Consequently, the application of the Alloy language contributes to a higher level of confidence in the validity of ecosystem requirements.

Moreover, to verify the properties and constraints of a given model, the Alloy Analyzer, a finite model finder that accompanies the Alloy language, can be used to automatically check, within a finite bound, whether an Alloy model satisfies the specified constraints \cite{torlakKodkodRelationalModel2007}. Alloy is used for a variety of purposes including the precise specification of complex systems. In this context, the automated analysis capabilities can help identify potential errors or design flaws during the verification phase.

In this section, we present a translation from UML to Alloy. Then we define the semantics of the core concepts in the Alloy language followed by a formalization of our metamodel, presented in Section \ref{section-runningexample}. Moreover, we present the CRESTO Alloy instance and finally, we present the verification of safety and liveness properties, with respect to CRESTO.

\subsection{Translation from UML to Alloy} \label{sect-umltoalloy}

Semantics, in the context of formal languages, refers to the meaning or interpretation of the symbols and expressions used in a language \cite{scottLogicProgrammingLanguages1977}. Specifically, we focus on the operational semantics, which defines the meaning of a formal language by describing the execution and evaluation of its expressions in terms of state transitions \cite{plotkinStructuralApproachOperational} to establish a mapping between UML and Alloy expressions. Hence, we can harness the complementary strengths of both languages, enabling a comprehensive and precise representation of complex systems.

The remainder of this sub-section presents a manual mapping between UML and Alloy, which we performed in the context of this article, highlighting the correspondences between their respective concepts, such as classes and signatures, attributes and fields, relationships and cardinalities, and inheritance.

\begin{itemize}
\item Classes and Signatures:
\begin{itemize}
\item UML classes are mapped to Alloy signatures, enabling the definition of objects structures in each respective language.
\item UML abstract classes are mapped to Alloy abstract signatures, which provide a base template that other classes or signatures can extend.
\end{itemize}
\item Attributes and Fields:
\begin{itemize}
    \item UML class attributes are mapped to Alloy signature fields, which represent the properties or attributes of objects in the corresponding set.
    \item UML attribute data types are mapped to corresponding Alloy data types such as Int or String, which constrain the values that can be assigned to a field.
\end{itemize}

\item Relationships and Cardinalities:
\begin{itemize}
    \item UML relationships such as association, aggregation and composition are mapped to Alloy signature fields with appropriate multiplicity, which specify the number of related objects allowed in each direction.
    \item UML relationship cardinalities are expressed using Alloy set cardinalities such as one, lone, some, and set, which specify the number of objects that can be related to another object through a particular relationship.
 \end{itemize}

\item Inheritance:
\begin{itemize}
    \item UML inheritance is mapped to Alloy's `extends' keyword, which allows a signature to inherit the properties and relationships of another signature.
\end{itemize}
\end{itemize}

\subsection{Formal definition of Alloy core concepts} \label{sect-formalalloy}

In this sub-section, we provide a mathematical formal definition and description of the semantics of the Alloy language, using first-order logic. Within the scope of this paper, we consider the Alloy specification only as a collection of signatures, facts, and predicates. It is worth mentioning that the Alloy language also includes other concepts, such as assertions, functions, and modules, which contribute to its semantics. Our aim in this section is to define the core principles and underlying characteristics of the Alloy language that are relevant in our formalization of ecosystem restoration requirements verification in the context of formal methods and software engineering.

\begin{definition}
	An Alloy specification, denoted as $spec$, is defined as a triple $\langle$Sig, Fact, Pred$\rangle$ where:
	\begin{itemize}
		\item $Sig$ is a set of Alloy signatures.
		\item $Fact$ is a set of Alloy facts.
		\item $Pred$ is a set of Alloy predicates.
	\end{itemize}
\end{definition}

\begin{definition}
	An Alloy signature $s \in Sig$ is defined as a triple $\langle$Atoms, Fields, Constraints$\rangle$ where:
	\begin{itemize}
		\item Atoms $\subseteq ATOMS$, where $ATOMS$ is the set of all possible atoms in the universe.
		\begin{itemize}
		\item $\forall \mathrm{a} \in \mathrm{Atoms}$, $a$ belongs to exactly one signature. Each atom $a$ can be thought of as an element of a signature $s$.
		\end{itemize}
		\item Fields consists of a relation $R \subseteq Sig \times Types$, where:
		\begin{itemize}
			\item $Types$ is a set of types, which can be either user signature \footnote{User signatures are explicitly defined by the modeler within the Alloy specification to represent specific concepts or relationships part of the problem domain being modeled.} or built-in signatures \footnote{Built-in signatures are predefined in the Alloy language and represent fundamental data types to use across any Alloy model.} of the Alloy language.
			\item Each field represents a relationship between instances of $s$ and other signatures (user or built-in), with varying cardinality based on the model's definition.
		\end{itemize}
		
		\item Constraints comprise a set of constraints, denoted by a set of first-order logical formulas, $\varphi$.
		\begin{itemize}
			\item Each constraint $\varphi$ is a logical expression involving Atoms and Fields of the signature $s$.
		\end{itemize}

	\end{itemize}
\end{definition}

\begin{definition}
An Alloy fact $f \in Fact$ is defined as a constraint that is applied to $spec$, represented as a first-order logical formula $\phi$:
\begin{itemize}
    \item Each fact $f$ is a logical expression involving Atoms, Fields, and Signatures present in the Alloy specification $spec$.
\end{itemize}

\end{definition}

\begin{definition}
An Alloy predicate $p \in Pred$ is defined as a reusable constraint represented as a parameterized first-order logical formula $\xi$:
\begin{itemize}
    \item Each predicate $p$ has a set of parameters and is a logical expression involving Atoms, Fields, Signatures as well as the parameters.
    \item Predicates can be invoked within facts, assertions, or other predicates, providing a modular way to express constraints and properties within $spec$.
\end{itemize}

\end{definition}

\subsection{Metamodel formalization}

Following the general translational rules of the Sub-section \ref{sect-umltoalloy} and the formal deﬁnition of the semantics of the Alloy language using first-order logic in Sub-section \ref{sect-formalalloy}, we now present a detailed Alloy formalization of our semi-formal UML metamodel, previously presented in Section \ref{req-section}. The formalized metamodel serves as the foundation for specifying ecosystem restoration requirements. The formalization of the metamodel concepts are systematically presented using signatures, fields, and constraints, combined with their Alloy implementation, which constitutes the metamodel specification.

\paragraph{\textbf{Ecosystem}} is defined formally as a signature $s_{Ecosystem} \in Sig$, where:

\begin{itemize}
\item Fields:
\begin{itemize}
\item $R_{hasSpecies} \subseteq s_{Ecosystem} \times s_{Species}$
\item $R_{regions} \subseteq s_{Ecosystem} \times s_{Region}$
\item $R_{restorationGoal} \subseteq s_{Ecosystem} \times s_{RestorationGoal}$
\item $R_{subEcosystems} \subseteq s_{Ecosystem} \times s_{Ecosystem}$
\end{itemize}

\item Constraints:
\begin{itemize}
\item $\varphi_{Ecosystem} = { \forall e \in s_{Ecosystem} : |R_{regions}(e)| \ge 1 }$
\item $\varphi_{Ecosystem} = { \forall e \in s_{Ecosystem} : |R_{subEcosystems}(e)| \ge 0 }$
\end{itemize}
\end{itemize}
The constraints on the Ecosystem signature indicate that each Ecosystem must have at least one associated Region and can have zero or more sub-ecosystems. The Alloy code representation of the Ecosystem concept is shown below:

\begin{lstlisting}[language=alloy, caption={Alloy code for the Ecosystem concept.}, label={lst:ecosystem}]
sig Ecosystem {
	hasSpecies: set Species,
	regions: set Region,
	restorationGoal: lone RestorationGoal,
	subEcosystems: set Ecosystem
} {
	#regions >= 1
	#subEcosystems >= 0
}
\end{lstlisting}

In Listing \ref{lst:ecosystem}, we define the Ecosystem signature with four fields: hasSpecies, regions, restorationGoal, and subEcosystems, representing sets of Species, Region, a single RestorationGoal, and a set of Ecosystem objects, respectively.

\paragraph{\textbf{Region}} is defined formally as a signature $s_{Region} \in Sig$, where:

\begin{itemize}
\item Fields:
\begin{itemize}
\item $R_{lines} \subseteq s_{Region} \times s_{Segment}$
\end{itemize}

\item Constraints:
\begin{itemize}
\item $\varphi_{Region} = { \forall r \in s_{Region} : |R_{lines}(r)| \ge 3 }$
\end{itemize}
\end{itemize}
The constraint on the Region signature states that each Region must have at least three associated Segments. The constraint is based on the fact that a region is a two-dimensional object that requires a closed boundary to enclose an area. To form a closed boundary, at least three line segments are needed that meet at three distinct points to create a closed shape, such as a triangle. Therefore, the constraint ensures that each Region object has at least three associated Segments to form a closed shape, which is necessary to define a region. The Alloy code representation of the Region concept is shown below:

\begin{lstlisting}[language=alloy, caption={Alloy code for the Region concept.}, label={lst:region}]
sig Region {
	lines: set Segment
} {
	#lines >= 3
}
\end{lstlisting}

In Listing \ref{lst:region}, we define the Region signature with one field named lines, corresponding to a set of Segment. The cardinality constraint, as previously mentioned, ensures that each Region has at least three associated Segments.

\paragraph{\textbf{Segment}} is defined formally as a signature $s_{Segment} \in Sig$, where:

\begin{itemize}
\item Fields:
\begin{itemize}
\item $R_{startX} \subseteq s_{Segment} \times String$
\item $R_{startY} \subseteq s_{Segment} \times String$
\item $R_{endX} \subseteq s_{Segment} \times String$
\item $R_{endY} \subseteq s_{Segment} \times String$
\end{itemize}

\item Constraints: None
\end{itemize}
The Segment signature does not have any constraints, as it simply represents a line segment defined by two endpoints with x and y coordinates, simplified as a String type. The Alloy code representation of the Segment concept is shown below:

\begin{lstlisting}[language=alloy, caption={Alloy code for the Segment concept.}, label={lst:segment}]
sig Segment {
	startX: one String,
	startY: one String,
	endX: one String,
	endY: one String
}
\end{lstlisting}

In Listing \ref{lst:segment}, we define the Segment signature with four fields: startX, startY, endX, and endY, representing the x and y coordinates of the start and end points of a line segment.

\paragraph{\textbf{Species}} is defined formally as an abstract signature $s_{Species} \in Sig$, where:

\begin{itemize}
\item Fields:
\begin{itemize}
\item $R_{inhabits} \subseteq s_{Species} \times s_{Ecosystem}$
\item $R_{scientificName} \subseteq s_{Species} \times String$
\end{itemize}

\item Constraints:
\begin{itemize}
\item $\varphi_{Species} = { \forall sp \in s_{Species} : |R_{scientificName}(sp)| = 1 }$
\end{itemize}
\end{itemize}
The constraint on the Species signature indicate that each Species must have exactly one associated scientific name. The Alloy code representation of the Species concept is shown below:

\begin{lstlisting}[language=alloy, caption={Alloy code for the Species concept.}, label={lst:species}]
abstract sig Species {
	inhabits: set Ecosystem,
	scientificName: one String,
}
\end{lstlisting}

In Listing \ref{lst:species}, we define the abstract Species signature with two fields: inhabits and scientificName, representing a set of Ecosystem and a String, respectively.

\paragraph{\textbf{Property}} is defined formally as a signature $s_{Property} \in Sig$, where:

\begin{itemize}
\item Fields:
\begin{itemize}
\item $R_{origin} \subseteq s_{Property} \times String$
\item $R_{originType} \subseteq s_{Property} \times s_{PropertyType}$
\item $R_{weight} \subseteq s_{Property} \times Int$
\item $R_{usedInRestoration} \subseteq s_{Property} \times s_{RestorationGoal}$
\end{itemize}

\end{itemize}
The Property signature does not have any constraints, as it represents a characteristic of the ecosystem restoration goal that is derived from a specific origin and has a certain weight. The Alloy code representation of the Property concept is shown below:

\begin{lstlisting}[language=alloy, caption={Alloy code for the Property concept.}, label={lst:property}]
sig Property {
	origin: one String,
	originType: one PropertyType,
	weight: one Int,
	usedInRestoration: set RestorationGoal
}
\end{lstlisting}

In Listing \ref{lst:property}, we define the Property signature with four fields: origin, originType, weight, and usedInRestoration, representing a String, a single PropertyType, an integer value, and a set of RestorationGoal, respectively.

\paragraph{\textbf{RestorationGoal}} is defined formally as a signature $s_{RestorationGoal} \in Sig$, where:

\begin{itemize}
\item Fields:
\begin{itemize}
\item $R_{on} \subseteq s_{RestorationGoal} \times s_{Ecosystem}$
\item $R_{restorationProp} \subseteq s_{RestorationGoal} \times s_{Property}$
\item $R_{goalType} \subseteq s_{RestorationGoal} \times s_{GoalType}$
\end{itemize}

\item Constraints:
\begin{itemize}
\item $\varphi_{RestorationGoal} = { \forall rg \in s_{RestorationGoal} : |R_{on}(rg)| = 1 }$
\item $\varphi_{RestorationGoal} = { \forall rg \in s_{RestorationGoal} : |R_{restorationProp}(rg)| = 1 }$
\item $\varphi_{RestorationGoal} = { \forall rg \in s_{RestorationGoal} : |R_{goalType}(rg)| \ge 1 }$
\end{itemize}
\end{itemize}
The constraints on the RestorationGoal signature indicate that each instance must be associated with exactly one Ecosystem, exactly one Property, and at least one GoalType. The Alloy code representation of the RestorationGoal concept is shown below:

\begin{lstlisting}[language=alloy, caption={Alloy code for the RestorationGoal concept.}, label={lst:restorationgoal}]
sig RestorationGoal {
	on: one Ecosystem,
	restorationProp: one Property,
	goalType: some GoalType,
} {
	#goalType >= 1
}
\end{lstlisting}

In Listing \ref{lst:restorationgoal}, we define the RestorationGoal signature with three fields: on, restorationProp, and goalType, representing a single Ecosystem, a single Property, and a set of GoalType, respectively.

\paragraph{\textbf{PropertyType}} is defined formally as an abstract signature $s_{PropertyType} \in Sig$ without any fields or constraints. It serves as a base for more specific types of properties (such as "Human" or "Natural") related to the ecosystem restoration process. The Alloy code representation of the PropertyType concept is shown in Listing \ref{lst:propertytype}.

\begin{lstlisting}[language=alloy, caption={Alloy code for the PropertyType concept.}, label={lst:propertytype}]
abstract sig PropertyType {}
\end{lstlisting}

\paragraph{\textbf{GoalType}} is defined formally as an abstract signature $s_{GoalType} \in Sig$, without any fields or constraints. It serves as a base for more specific types of goals related to the ecosystem restoration process. In Listing \ref{lst:goaltype}, we define the abstract GoalType signature.

\begin{lstlisting}[language=alloy, caption={Alloy code for the GoalType concept.}, label={lst:goaltype}]
abstract sig GoalType {}
\end{lstlisting}

\subsection{Formalization of CRESTO} \label{sect-crestoform}

In the previous sub-section, we presented the detailed Alloy formalization of our semi-formal UML metamodel, established through the application of the translational rules outlined in sub-section \ref{sect-umltoalloy} and the formal definition of the semantics of the Alloy language using first-order logic in sub-section \ref{sect-formalalloy}. 

Building on this foundation, in this sub-section, we apply the concepts and constructs derived from the formalized metamodel to the CRESTO running example. Our objective is to illustrate how our formalized metamodel can be used in the context of a real-world ecosystem restoration project, highlighting the process of specifying requirements using a formal approach.

\begin{lstlisting}[language=alloy, caption={Alloy instances for the CRESTO running example.}, label={lst:cresto-instances}]
// Ecosystem instances
one sig GCA extends Ecosystem {}
one sig ModuloII, ControlZone extends Ecosystem {}

fact {
	GCA.subEcosystems = ModuloII + ControlZone
}

// Region instances
one sig ModuloRegion, ControlRegion extends Region {}

// Segment instances
one sig S1, S2, S3, S4, S5, S6, S7, S8 extends Segment {}

fact {
	ModuloRegion.lines = S1 + S2 + S3 + S4
	ControlRegion.lines = S5 + S6 + S7 + S8
}

// Species instances
one sig Hyparrhenia, Byrsonima extends Species {}

fact {
	Hyparrhenia.inhabits = ModuloII
	Byrsonima.inhabits = ModuloII
	Hyparrhenia.scientificName = "Hyparrhenia rufa"
	Byrsonima.scientificName = "Byrsonima crassifolia"
}

// Property instances
one sig OrangeWaste extends Property {}
fact {
	OrangeWaste.origin = "Juice Factory"
	OrangeWaste.originType = Human
	OrangeWaste.weight = 1
	OrangeWaste.usedInRestoration = ForestRestoration
}

// RestorationGoal instances
one sig ForestRestoration extends RestorationGoal {}
fact {
	ForestRestoration.on = ModuloII
	ForestRestoration.restorationProp = OrangeWaste
	ForestRestoration.goalType = Biodiversity + CarbonSequestration + SoilImprovement
}

// PropertyType instances
one sig Human extends PropertyType {}

// GoalType instances
one sig Biodiversity, CarbonSequestration, SoilImprovement extends GoalType {}
\end{lstlisting}
We now present a detailed explanation of the Alloy instance for the CRESTO running example, as shown in Listing \ref{lst:cresto-instances}.

\begin{itemize}
	\item Ecosystem instances: We define GCA, ModuloII, and ControlZone as instances of the Ecosystem. The fact specifies that GCA has two sub-ecosystems: ModuloII and ControlZone.
	\item Region instances: We define ModuloRegion and ControlRegion as instances of the Region.
	\item Segment instances: We define eight instances of the Segment: S1, S2, S3, S4, S5, S6, S7, and S8. The fact associates the segments with the regions; ModuloRegion contains segments S1 to S4, while ControlRegion contains segments S5 to S8.
	\item Species instances: We define Hyparrhenia and Byrsonima as instances of the Species. The fact specifies that both species inhabit the ModuloII ecosystem and sets their scientific names.
	\item Property instances: We define OrangeWaste as an instance of the Property. The fact sets the origin, originType, weight, and restoration usage of OrangeWaste.
	\item RestorationGoal instances: We define ForestRestoration as an instance of the RestorationGoal. The fact associates ForestRestoration with the ModuloII ecosystem, OrangeWaste property, and sets the goal types: Biodiversity, CarbonSequestration, and SoilImprovement.
	\item PropertyType instances: We define Human as an instance of PropertyType. This represents the origin type of the OrangeWaste property, indicating that it originates from human activities.
	\item GoalType instances: We define Biodiversity, CarbonSequestration, and SoilImprovement as instances of GoalType. These instances represent the types of restoration goals in CRESTO.
\end{itemize}
This presentation shows how the various instances relate to each other, providing a comprehensive representation of the ecosystem restoration requirements with respect to CRESTO.

\subsection{Verification of Safety and Liveness Properties}

Building upon the previous sections, where we formally defined the core concepts of Alloy and presented the formalization of our UML metamodel, we now focus on verifying safety and liveness properties in our Alloy specification model for natural ecosystem requirements. Safety properties ensure that undesirable situations do not occur throughout the system's operation, whereas liveness properties guarantee that the system progresses towards desired outcomes, such to a restoration goal in this context. By verifying these properties, we aim to identify potential issues or inconsistencies in our specification model, thereby enhancing its reliability and robustness.

Thereby, we utilize the Alloy Analyzer to assess the validity of our model, presented in sub-section \ref{sect-crestoform}, against a short list of safety and liveness properties, as a proof of concept of our approach. The subsequent subsections outline the derivation of these properties, the associated Alloy assertions, and the analysis of the verification results obtained using the Alloy Analyzer.

\subsubsection{Safety Properties}

\begin{figure}[htp]
\centering
     \includegraphics[angle=90, width=0.9\textwidth]{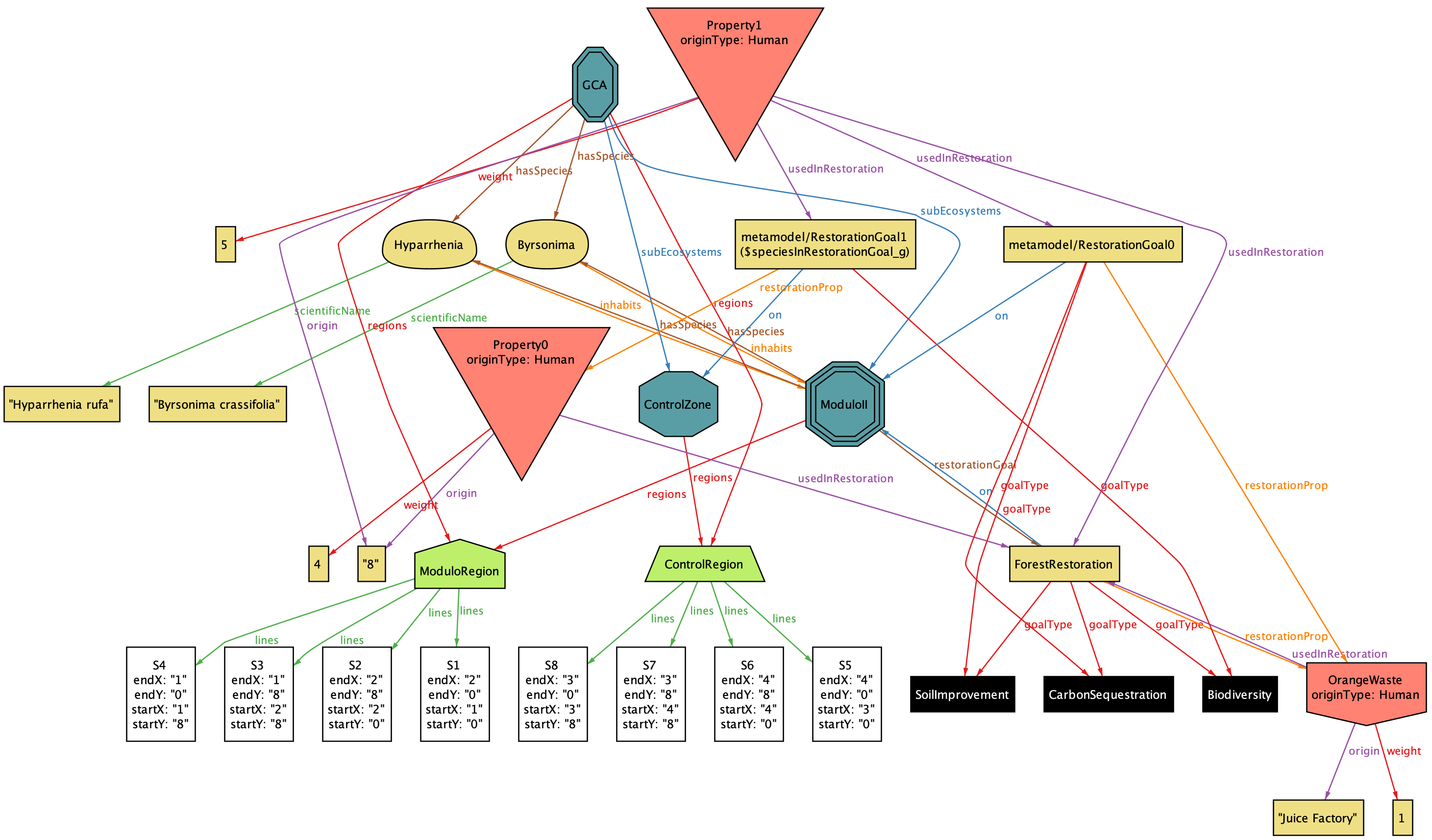}
      \caption{Alloy Analyzer output presenting a counterexample for the "speciesInRestorationGoal" safety property check. The counterexample demonstrates that species may not be present in an ecosystem with a restoration goal associated with it, which invalidates the defined safety property in our formal model.}
	     \label{fig:DistinctRegionsHaveDistinctSegments}
\end{figure}

\begin{enumerate}

\item \textit{Species required for a restoration goal are already present within the ecosystem being restored.}
This safety property is important for ecosystem restoration verification for several reasons. First, ensuring that the species required for a restoration goal are present within the ecosystem being restored helps maintain the consistency of the restoration process. This consistency is vital for successful restoration projects, as it ensures that the targeted ecosystem has the necessary species to meet the desired restoration objectives. Second, if the required species are not present within the ecosystem, the restoration process may face various challenges. For example, the absence of  some species could result in a slower ecosystem restoration, as presented in CRESTO. Lastly, by verifying the presence of the required species in the target ecosystem, stakeholders can better allocate resources for the restoration process, allowing the prioritization of other essential aspects of the restoration project. In Figure \ref{fig:DistinctRegionsHaveDistinctSegments}, we show the Alloy Analyzer output presenting a counterexample for the safety property check. Such counterexample allows stakeholders using our approach to detect inconsistencies, iteratively refine their model, and enhance their restoration process's accuracy and robustness.

\begin{lstlisting}[language=alloy]
assert speciesInRestorationGoal {
	all g: RestorationGoal | g.on.hasSpecies = (g.restorationProp.usedInRestoration.on.hasSpecies)
}
check speciesInRestorationGoal
\end{lstlisting}

\item \textit{The restoration process should operate on an ecosystem.}
This safety property emphasizes the importance of ensuring that the restoration goal operates specifically on an ecosystem, rather than on an unrelated or irrelevant entity. This is essential to guarantee that the formal verification and restoration requirements are applied to the appropriate context. In Figure \ref{fig:ProcessShallOperateOnEcosystem-counterexample}, the Alloy Analyzer did not find counterexamples of the property. This validity result suggests that our formalization effectively ensures that the restoration process is applied to the appropriate context, specifically targeting ecosystems and contributing to their restoration goal.
\begin{lstlisting}[language=alloy]
assert RestorationShallOccurOnEcosystem {
    all rp: RestorationGoal | rp.on in Ecosystem
}
check RestorationShallOccurOnEcosystem
\end{lstlisting}
\end{enumerate}

\subsubsection{Liveness Properties}

\begin{figure}[htp]
\centering
     \includegraphics[width=0.85\textwidth]{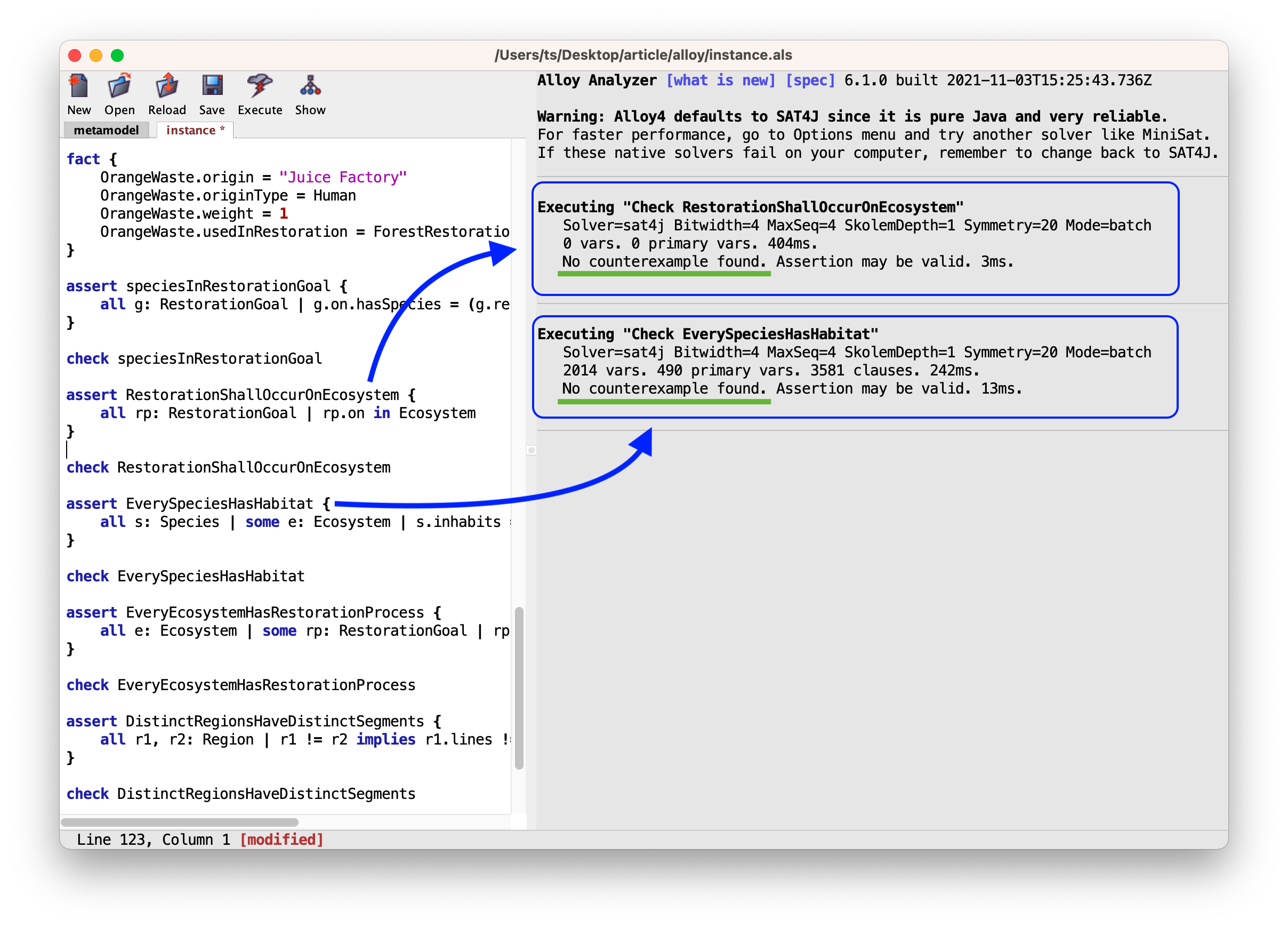}
      \caption{Alloy Analyzer output for the "RestorationShallOccurOnEcosystem" safety property and "EverySpeciesHasHabitat" liveness property. The output reveals that no counterexamples are found for both properties, indicating that the assertion may be valid within our formal model.}
	     \label{fig:ProcessShallOperateOnEcosystem-counterexample}
\end{figure}

\begin{enumerate}
\item \textit{Eventually, every species should have a habitat.}
This liveness property highlights the important goal of ecosystem restoration efforts: providing suitable habitats for every species. Ensuring that each species has a habitat is essential to support biodiversity and maintain the overall health of the ecosystem. By emphasizing this property, the formal verification process can help guide restoration efforts towards achieving this critical objective, contributing to the long-term success of ecosystem recovery. In our verification, depicted in Figure \ref{fig:ProcessShallOperateOnEcosystem-counterexample}, we were able to confirm the validity of the property.
	\begin{lstlisting}[language=alloy]
assert EverySpeciesHasHabitat {
    all s: Species | some e: Ecosystem | s.inhabits = e
}
check EverySpeciesHasHabitat
\end{lstlisting}
	\item \textit{Eventually, every ecosystem has a restoration goal associated with it.}
This liveness property is essential to create a more resilient and sustainable global environment, as it encourages the development and implementation of restoration plans for all ecosystems, regardless of their current state. In contrast to the previous liveness property, our model execution with the Alloy Analyzer identified a counterexample. The presence of such counterexample indicates that our formalization approach can effectively allow the detection of important cases that are essential to uncover during verification phases.
	\begin{lstlisting}[language=alloy]
assert EveryEcosystemHasRestorationProcess {
    all e: Ecosystem | some rp: RestorationGoal | rp.on = e
}
check EveryEcosystemHasRestorationProcess

\end{lstlisting}
\end{enumerate}

\section{Assessment}
The following assessment provides a summary of the research method used, as well as a factual overview of the strengths and weaknesses of our approach and highlights key contributions in the context of software engineering and formal methods.

Our research method, which is based on a case study, allows us to explore the specific context of ecosystem restoration requirements and obtain valuable insights into our formalization approach. By analyzing a single, simple case in detail, we can uncover the nuances and intricacies involved in the verification of ecosystem restoration requirements. This detailed understanding of the case, combined with our formalization in Alloy, enables us to develop a more general and abstract solution that can be applied to a wider range of cases in the domain of ecosystem restoration requirements. The use of a case study research method allows us to establish a solid foundation in real-world contexts and effectively address the practical challenges faced by stakeholders in the field of ecosystem restoration requirements. However, its weakness is related to the lack of generalizability. By focusing on a single case study, we may not be able to thoroughly evaluate the applicability and effectiveness of our approach across a broader range of case studies differing from the initial one. Although case studies can provide valuable insights into specific instances, it may be difficult to generalize our approach to other cases, potentially limiting the overall robustness and validity of our approach.

In the context of ecosystem restoration, the application of Alloy and UML offers significant benefits, including clarity in defining requirements, rigorous verification capabilities and early error detection. However, these languages also present weaknesses such as complexity and scalability, particularly for Alloy.  Alloy, as any formal language can be complex and may require a significant learning curve, posing a barrier for stakeholders who are not familiar with these languages. Scalability is crucial for any formal verification approach, as it determines the method's ability to handle larger, more complex models and requirements. While Alloy enables automatic analysis and provides an adaptable solution for various complexities of ecosystem requirements verification, it faces challenges when dealing with larger, more complex models. The Alloy Analyzer's finite model finder explores a vast solution space within a finite bound, based on the small scope hypothesis. This hypothesis asserts that most errors in a model can be detected within a small instance, allowing for efficient verification of complex requirements. The Alloy Analyzer leverages this principle, however, its performance may degrade as the complexity of the model increases \cite{jacksonSoftwareAbstractionsLogic2006}, which is a limitation to consider.

Nevertheless, our approach demonstrates its effectiveness by identifying inconsistencies and errors in the early stages of the software development lifecycle. By providing a formalized metamodel in Alloy, our method enables stakeholders to precisely capture the intended requirements of an ecosystem as well as restoration goals. This level of precision reduces ambiguities and enhances the clarity of communication among stakeholders, contributing to a higher level of confidence in the validity of ecosystem requirements.

Our work contributes to integrate formal methods into software engineering by offering a robust and efficient approach for formalizing and verifying ecosystem restoration requirements using Alloy. This enhances the quality and reliability of software systems while advancing formal methods and software engineering in this domain, laying the groundwork for future research and innovation.
 
\section{Related Work}
The application of modeling languages and formal methods has become increasingly important across various domains. In this related work section, we provide an overview of existing research utilizing UML and formal methods, highlighting the unique contributions of our approach in addressing the challenges of ecosystem restoration.

UML has been praised as the "de facto standard" of software engineering \cite{petreUMLPractice2013}. Despite this, its application extends beyond software engineering and into fields such as environmental modeling and sustainable ecosystem management \cite{courtneyDecisionSupportSystems2013}. For instance in \cite{khaiterConceptualizingEnvironmentalSoftware2018}, Khaiter et al. conceptualized a UML metamodel for an environmental software modeling framework, which serves as a tool for addressing sustainability tasks. Their work outlines the framework's multi-layered architecture and its primary software components, using UML diagrams to depict the internal functional logic of each component. Compared to those studies, our approach applies a semi-formal UML metamodel specifically for ecosystem restoration requirements and formalizes it with formal methods such as Alloy. Hence, our approach leverages the strengths of both approaches to create more robust and reliable models, particularly in the context of software engineering.

The use of formal methods remains relatively limited in the ecosystem domain, likely due to the complex, dynamic, and often non-linear nature of ecological systems that pose challenges for applying formal techniques. Nevertheless, there are a handful of studies where formal methods have been successfully employed within the ecosystem domain. 

For instance, in \cite{thomasModelcheckingEcologicalStatetransition2022}, Thomas et al. discuss the application of model-checking within ecosystems, automatically assessing the dynamics of ecological systems by verifying whether a state-transition graph satisfies a dynamical property expressed as a temporal logic formula. This article offers an inventory of existing ecological state-transition graphs and a clear presentation of the model-checking methodology, using ecosystem vegetation models as examples. Additionally, in \cite{konigsbergModellingVerificationAnalysis2017}, Konigsberg et al. present a formal methodology for modeling and verifying predator-prey interactions in an ecosystem using first-order logic and qualitative methods to verify satisfiability and address performance concerns. Finally, in \cite{largouetUseTimedAutomata2012}, Largouët et al. propose a qualitative modeling approach based on timed automata formalism, coupled with model-checking techniques, to evaluates ecosystem property dynamics and temporal evolution in response to various management options.

In contrast to the cited papers, our work sets itself apart by adopting a comprehensive approach to ecosystem restoration and by formalizing the restoration requirements using the Alloy language. Our key contribution lies in proposing a rigorous approach for specifying and verifying ecosystem restoration requirements, encompassing a wider ecological context and addressing various aspects of the restoration process. 
\section{Perspectives}
This paper provides a foundation for further exploration in formal verification of ecosystem requirements within software engineering. Several directions for future work can build upon and refine the current work.

It is important to address the generalizability limitation of our research method. Hence, future work could involve conducting additional case studies or employing complementary research methods to further assess and validate the applicability of our approach to ecosystem restoration requirements verification in software engineering. Additionally, expanding the metamodel to allow the formalization and specification of more concepts related to ecosystems is important for improving the overall scope and effectiveness of our method.

Moreover, we are working on the application of the Alloy specification language for the formalization of a conceptual framework that delineates the concepts of dependability and resilience \cite{guelfiFormalFrameworkDependability2011}, which can be applied in the context of natural ecosystem restoration. This initiative aims at establishing a rigorous foundation for modeling the fundamental characteristics that contribute to the robustness and adaptability of ecological systems. Consequently, this will facilitate the development of more efficacious strategies for the restoration of deteriorated ecosystems and the assurance of their long-term sustainability. The motivation for this aspect of our research is to leverage formal verification with Alloy in order to connect the theoretical comprehension and practical execution of ecological restoration efforts, thus providing a more rigorous and reliable basis for decision-making. This approach aims at empowering decision-makers to make well-informed decisions that promote ecosystem health and resilience, ensuring that restoration efforts are grounded in a robust and dependable framework. 
\section{Conclusion}
In this paper, we introduced a formal approach for modeling and verifying the requirements of ecosystems restoration goals using UML and Alloy. By enhancing the rigor and reliability of requirements engineering and verification in the domain of natural ecosystems restoration, our approach offers valuable insights for researchers, practitioners, and policymakers alike. Furthermore, the verified ecosystem requirements model could serve as a foundation for AI-based and remote sensing applications in ecosystem restoration data management. Ultimately, this work contributes significantly to the field of formal software engineering applied to ecosystems restoration, helping to ensure the development of reliable ecosystem specifications that meet specified restoration goals. 
\bibliographystyle{ieeetr}
\bibliography{phd-tiago.bib,phd-tiago-overleaf,phd-tiago-additional-ref}

\section*{Authors}
\noindent {\bf Tiago Sousa} is pursuing his PhD in Computer Science at the University of Luxembourg. His research interests include Software Engineering \& Artificial Intelligence related to the domain of resilient ecosystems and remote sensing.\\

\noindent {\bf Dr. Benoît Ries} is a permanent research scientist in the Software engineering and Artificial intelligence Group on Ecosystem management (SAGE) in the Department of Computer Science of the University of Luxembourg. His current research interests focus in model-driven software engineering for AI-based systems, in particular focusing on the requirements and design phases applied to the domain of resilient natural ecosystems.\\

\noindent {\bf Dr. Nicolas Guelfi} is full professor at University of Luxembourg, since 1999. He is head of the SAGE group in the Department of Computer Science of the University of Luxembourg. \\

\end{document}